\documentclass[10pt,aps,prl,reprint,twocolumn, nofootinbib,aps,superscriptaddress]{revtex4-1}

\usepackage{amssymb,amsmath}
\usepackage{graphicx}
\usepackage{color}
\usepackage{listings}
\usepackage{float}
\usepackage{wrapfig}
\usepackage{physics}
\usepackage{amsmath}
\usepackage{gensymb}
\usepackage{soul}
\usepackage[format=plain]{caption}
\usepackage{subcaption}

\usepackage{upgreek}
\usepackage{enumerate} 
\usepackage[linkcolor = blue, citecolor = blue, urlcolor = blue, colorlinks = true]{hyperref}
\usepackage{commath,amssymb}
\usepackage{ushort}
\usepackage{multirow}
\usepackage[usenames,dvipsnames]{xcolor}
\usepackage{cleveref}
\usepackage{epstopdf}
\usepackage[exponent-product = \cdot]{siunitx}

\crefname{equation}{Eq.}{Eqs.}
\crefname{figure}{Fig.}{Figs.}
\crefrangeformat{equation}{Eqs.~#3(#1)#4--#5(#2)#6}

\hypersetup{
    colorlinks=true,
    linkcolor=blue,
    citecolor=red}

\begin{document}

\title{Slip length dependent propulsion speed of catalytic colloidal swimmers near walls}
\author{Stefania Ketzetzi}
\affiliation{Soft Matter Physics, Huygens-Kamerlingh Onnes Laboratory, Leiden University, P.O. Box 9504, 2300 RA Leiden, The Netherlands}
\author{Joost de Graaf}
\affiliation{Institute for Theoretical Physics, Center for Extreme Matter and Emergent Phenomena, Utrecht University, Princetonplein 5, 3584 CC Utrecht, The Netherlands}
\author{Rachel P. Doherty}
\affiliation{Soft Matter Physics, Huygens-Kamerlingh Onnes Laboratory, Leiden University, P.O. Box 9504, 2300 RA Leiden, The Netherlands}
\author{Daniela J. Kraft}
\affiliation{Soft Matter Physics, Huygens-Kamerlingh Onnes Laboratory, Leiden University, P.O. Box 9504, 2300 RA Leiden, The Netherlands}


\date{\today}

\begin{abstract}
Catalytic colloidal swimmers that propel due to self-generated fluid flows exhibit strong affinity for surfaces. We here report experimental measurements of significantly different velocities of such microswimmers in the vicinity of substrates made from different materials. We find that velocities scale with the solution contact angle $\theta$ on the substrate, which in turn relates to the associated hydrodynamic substrate slip length, as $V\propto(\cos\theta+1)^{-3/2}$. We show that such dependence can be attributed to osmotic coupling between swimmers and substrate. Our work points out that hydrodynamic slip at the wall, though often unconsidered, can significantly impact the self-propulsion of catalytic swimmers. 

\end{abstract}

\maketitle

Colloidal swimmers constitute a new class of non-equilibrium model systems, that also hold great promise for applications owing to their fast directed motion in liquid environments. A simple experimental realization of such microswimmers are spherical colloids half coated with Pt~\cite{Ebbens2018}. These colloids move autonomously in H$_2$O$_2$ solutions due to asymmetric catalytic reactions taking place on their surfaces~\cite{Golestanian2005} and are typically found self-propelling parallel to a substrate~\cite{Brown2014,Ke2010,Das2015,Simmchen2015}. This substrate-affinity leads to accumulation~\cite{Brown2014} and retention~\cite{Das2015,Simmchen2015,Brown2016} of swimmers at surfaces, such as walls and obstacles, and can be exploited as a means to guide their motion~\cite{Das2015,uspal2016guiding}. 

Strikingly, upon approaching a surface, numerical and theoretical models predict both an increase or decrease in swimming velocity depending on the considered propulsion mechanism and the physico-chemical properties of the swimmer and wall~\cite{Popescu2009,Crowdy2013,Chiang2014,Uspal2015, Ibrahim2015,Mozaffari2016, Shen2018}. Even more so, experimental observations also hint at non-negligible substrate effects for synthetic swimmers~\cite{Wei2018sub,Palacci2014sub}. The perhaps most puzzling observation for catalytic swimmers is the inconsistency in swimming velocities under comparable experimental conditions. For example, velocities as disparate as 3 $\mu$m/s~\cite{Howse2007} and 18 $\mu$m/s~\cite{Brown2014} were found for polystyrene spheres with 5 nm Pt coating in 10\% H$_2$O$_2$. This discrepancy is even more surprising when one considers that the slower speeds were observed for the smaller species, whereas the velocity of Pt-coated swimmers should scale inversely with size~\cite{Ebbens2012}. Resolving these discrepancies will facilitate the development of a quantitative framework for these microswimmers.

Recent experiments on bimetallic microswimmers moving along glass surfaces showed a velocity decrease upon functionalizing the glass with polyelectrolytes~\cite{Wei2018sub}. Surprisingly, the decrease on the velocity was observed for both positively and negatively charged polyelectrolytes, indicating that wall zeta potential does not have a dominant effect on the velocity of electrophoretic swimmers.
At the same time, the effect of two different substrates, glass and Au-coated glass, on photo-activated TiO$_2$/SiO$_2$ swimmers was considered~\cite{Holterhoff2018}. Interestingly, 4 $\mu$m/s average velocities were found on Au, as opposed to 3 $\mu$m/s on glass. It was proposed, based on zeta potential values for Au and glass at neutral pH conditions, that the increase in the velocity stemmed from the lower zeta potential of the Au surface. However, neutral conditions are likely not met in H$_2$O$_2$ solutions. Results obtained using Au-coated substrates are hard to interpret, because Au could in principle catalyze H$_2$O$_2$ decomposition and thus interfere with the propulsion reaction. To corroborate the intriguing prospect that differences in swimming speeds originate from the substrate, surfaces other than Au ought to be examined. Furthermore, to pinpoint the origin of potential velocity differences a quantitative approach is required. Understanding potential surface effects on colloid self-propulsion is essential not only for their use as model systems, but also for future applications that may require motion in complex environments comprising obstacles or confining walls~\cite{Bechinger2016}. 

\begin{figure*}[ht]
  \bigskip
  \centering
    \includegraphics[width=\textwidth]{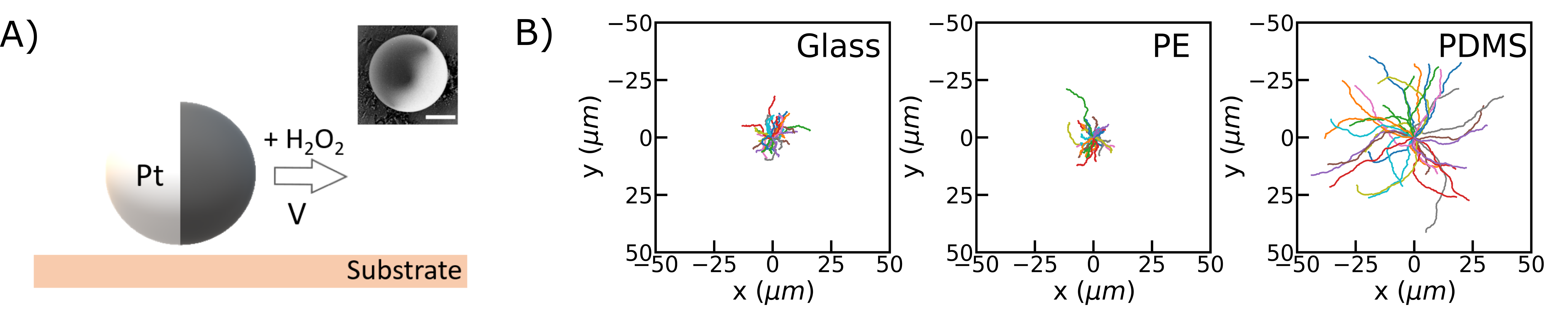}
 \caption{\textbf{Influence of the substrate on colloid self-propulsion.} A) Schematic of the experimental setup. The self-propulsion of 2.7 $\mu$m diameter Pt-coated colloids was observed in the same H$_2$O$_2$ solution and under fixed conditions on various substrates. All colloids were taken from the same batch. A Scanning Electron Microscope image of a representative colloid is shown in the inset, with the brighter hemisphere indicating the Pt coating. Scale bar is 1 $\mu$m. b) Typical 8 s active colloid trajectories on glass, polyethylene (PE) and  polydimethylsiloxane (PDMS).} 
 \label{Fig:substrates}
\end{figure*} 

In this Letter, we quantitatively examine the effect of various substrates, namely glass, glass coated with the organosilicon compound polydimethylsiloxane (PDMS), and plastic substrates made of a polyethylene (PE) or polystyrene (PS) derivative, on the velocity of catalytically self-propelled colloidal swimmers. 
Under otherwise fixed conditions, we observe significant differences in the velocities, which cannot be fully accounted for by the substrate zeta potential. 
Instead, we find that velocities upon different substrates fall on a single curve as a function of the solution-substrate contact angle which relates to the substrate slip length~\cite{BocquetAngle2008}. This finding indicates a velocity dependence on substrate slip for catalytic swimmers. After careful examination of the observed dependence in view of qualitative and scaling arguments, and accounting for possible couplings between swimmers and the substrate, we show that substrate-dependent velocities may result from osmotic coupling.

For all experiments, we used TPM colloids~\cite{Wel2017TPM} of diameter 2.7 $\mu$m half-coated with 4.9 nm of Pt by sputter-coating, see inset in Figure~\ref{Fig:substrates}A. Colloids were prepared in one batch, and thus any inhomogeneities arising from their preparation, including Pt thickness that affects H$_2$O$_2$ decomposition, should be universal. Measurements were taken with a 60x ELWD air objective (NA 0.7) on an inverted Nikon TI-E microscope at 18.92 fps in the dark typically within the hour after dispersing the colloids at dilute particle concentration ($\approx$ 10$^{-7}$ w/v) in deionized water containing 10\% H$_2$O$_2$. The colloids quickly reached the lower surface and continued to self-propel parallel to it, as illustrated in Figure~\ref{Fig:substrates}A. 
Figure~\ref{Fig:substrates}B shows 35 representative xy trajectories on glass, PE, and PDMS substrates acquired over a time interval of 8 s. 
We find that the colloids cover significantly greater distances on PDMS than on glass and PE, clearly demonstrating that the substrate affects colloid motion. To quantify the differences in the observed behavior, we first obtain the velocity of each individual colloid from its short-term mean squared displacement following Ref.~\cite{Howse2007}. We fit the corresponding probability density function (PDF) of the velocity with a log-normal distribution following Ref.~\cite{Goldstein2019} to obtain the velocity distribution parameters on each substrate. Details on the determination of colloid velocities can be found in SI Section I D. The most frequently encountered velocities, as obtained from the fitted peak position of each PDF, are 1.05~$\pm$ 0.09, 1~$\pm$ 0.2, and 2.8~$\pm$ 0.3 $\mu$m/s, above glass, PE, and PDMS, respectively. Interestingly, though all three substrates are chemically different, the colloids show similar velocities for two of the substrates and a notably different velocity for the third. In the absence of H$_2$O$_2$, however, the translational diffusion coefficients are similar, namely 0.099~$\pm$ 0.005, 0.098~$\pm$ 0.008, and 0.105~$\pm$ 0.005 $\mu$m$^2$/s, for glass, PE and PDMS, respectively. Therefore, substrate-dependent differences arise only in the active state. 

While velocities may be influenced by colloid properties, such as size~\cite{Ebbens2012}, roughness~\cite{Longbottom2018} and slip~\cite{Ajdari2006,Manjare2014}, these effects are negligible here since the same colloid batch was used in all experiments. Thus, the observed velocity differences arise from differences in the substrate properties. To quantitatively unravel the origin of our observations, we consider substrate properties that may influence colloid motion. The fluid flow generated by the anisotropic catalytic reaction on the swimmer surface~\cite{Campbell2018}, and hence the swimming velocity~\cite{Anderson1989}, has been predicted to be affected by the swimmer-wall distance~\cite{Popescu2009,Crowdy2013,Chiang2014,Uspal2015, Ibrahim2015,Mozaffari2016, Shen2018}, wall zeta potential~\cite{Chiang2014,Holterhoff2018} and wall surface inhomogeneities~\cite{Holterhoff2018}. Surprisingly, little consideration has been given until now on whether slip on the substrate impacts swimming velocities, even though slip on the colloid has already been shown to do so~\cite{Manjare2014}. Considering that hydrodynamic attraction in the active state pulls the colloids close to surfaces, to the extent that they even propel along the top of their container~\cite{Brown2014}, colloid-substrate distances are expected to be small. Pt-coated swimmers of 2.5 $\mu$m radius have been found to not swim over 200 nm steps~\cite{Simmchen2015}, and other experiments pointed out that distances may even be of the order of tens of nm~\cite{Takagi2013,Takagi2014}. Since wall slip lengths ranging from tens~\cite{Craig2001slip,Baudry2001slip, Bhushan2009,Bocquet2007} to hundreds~\cite{Pit2000slip,Bizonne2002slip} of nanometers and even micrometers~\cite{Zhu2001slip} have been reported, boundary conditions could strongly affect the velocity. Following Ref.~\cite{Ajdari2006}, we hypothesize that deviations from the no-slip condition on the substrate may enhance the nearby swimmer velocities.

\begin{figure}[!ht]
  \centering
    \includegraphics[width=0.4\textwidth]{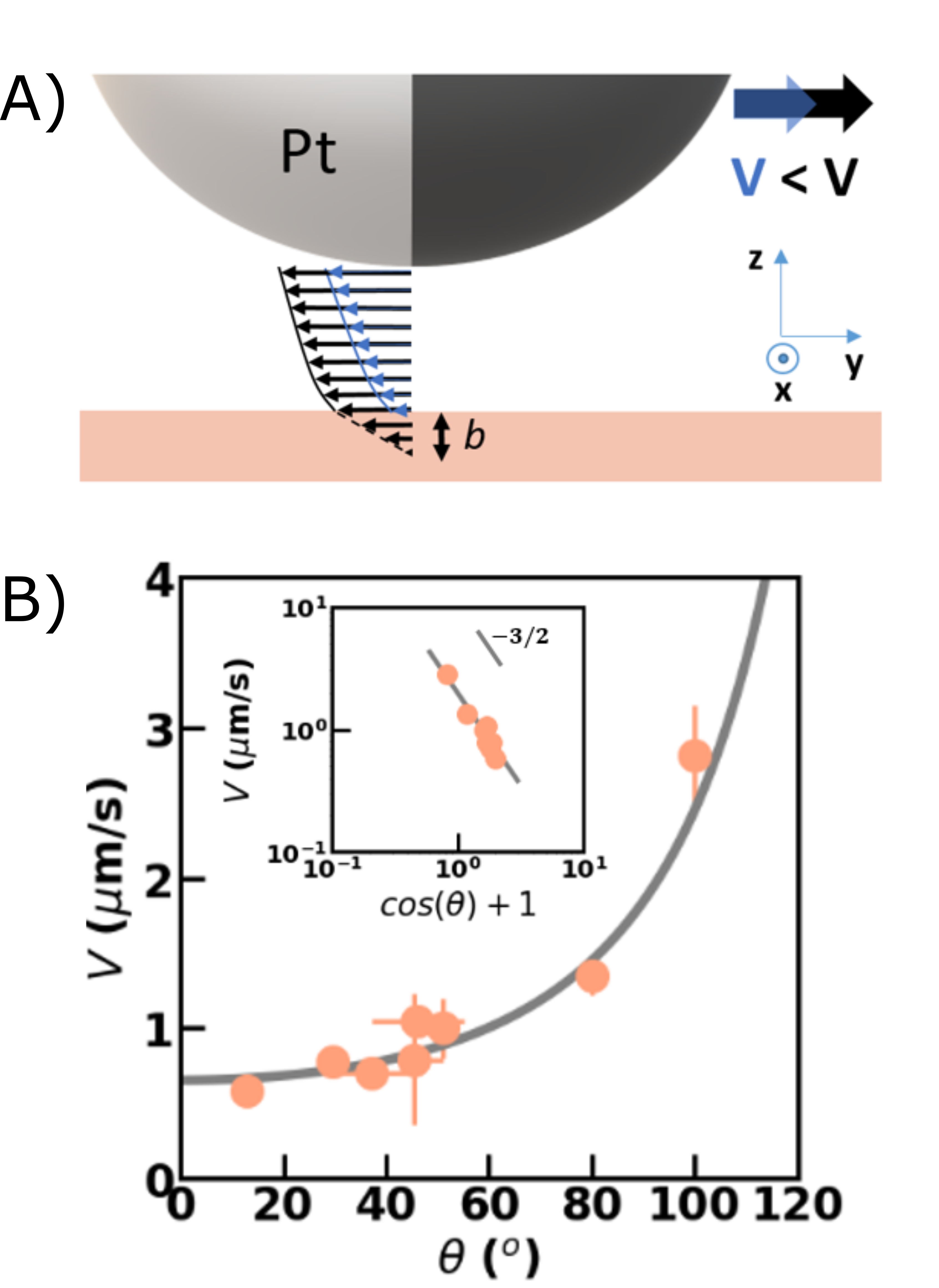}
\caption{\textbf{Slip dependence of the velocity of catalytic swimmers.} A) Schematic of the proposed model. At a given distance from the substrate, the swimming velocity $V$ resulting from the colloid-generated fluid flow is larger on a hydrophobic substrate due to the larger slip length $b$. Here, only the fluid flow velocity profile due to hydrodynamic slip on the wall is illustrated. B) Swimming velocity $V$ as a function of the contact angle $\theta$ and least squares fit (solid line) with $V = A (\cos{\theta}+1)^{-3/2}$ that follows from our model, with $A$ 1.84 $\mu$m/s. The inset shows the data on a log-log scale.}
\label{Fig:mechanism}
\end{figure}

Surface slip relates to liquid-solid interactions, and thus surface wetting properties, and generally, though not always, increases with increasing hydrophobicity and thus contact angle $\theta$~\cite{Bocquet2007, BocquetAngle2008, Bhushan2009, Siboulet2013}. Since a larger slip length $b$ on hydrophobic surfaces leads to a larger slip fluid velocity~\cite{kezirian1992hydrodynamics}, we hypothesize that this leads to a larger swimming velocity. Conversely, the no-slip approximation on hydrophilic surfaces would lead to a lower velocity, see Figure~\ref{Fig:mechanism}A. Indeed, the measured contact angles for the H$_2$O$_2$ solution agree with this hypothesis: $\theta$  

is 46~$\pm$ 9$\degree$, 51~$\pm$ 3$\degree$, and 100~$\pm$ 3$\degree$, for glass, PE, and PDMS, respectively. PE is normally hydrophobic, thus a modification had been performed by the supplier.

To further test this hypothesis, we modulated the hydrophilicity of the employed substrates and repeated the experiments. We increased the hydrophilicity of glass by either a cleaning procedure ($\theta= 29.5 \pm 3 \degree$) or treatment with HCl ($\theta= 13\pm 3\degree$) and observed a concomitant velocity decrease by 30\% and 45\% (0.78 $\pm$ 0.03 and 0.6 $\pm$ 0.05 $\mu$m/s), respectively. Conversely, when we rendered the glass more hydrophobic ($\theta= 80\pm 2$\degree), we found that the velocity increased by 28\% (1.35 $\pm$ 0.13 $\mu$m/s) compared to untreated glass. Similar behavior was seen on PDMS that was rendered hydrophilic through UV-ozone treatment ($\theta= 37 \pm 7 \degree$): colloidal particles propelled 4 times slower (0.7 $\pm$ 0.05 $\mu$m/s) than on the hydrophobic PDMS. Finally, we employed commercially available PS substrates that were rendered hydrophilic ($\theta= 46 \pm 6 \degree$) and found 0.8 $\pm$ 0.45 $\mu$m/s. We summarize these findings by plotting velocity as a function of $\theta$ in Figure \ref{Fig:mechanism}B. The collapse of the data onto a single curve suggests that $\theta$, and thus slip length, is the most relevant parameter and that other differences among substrates, besides their effect on $\theta$, are of lesser importance.

Next, we develop a quantitative framework for the slip-dependent swimming velocities. For our analysis we consider that the height above the substrate remains relatively unaffected by the change of substrate, as supported by our experimental measurements of the diffusion coefficient, see SI Section I D. When the height is left unperturbed by varying $\cos\theta$ --- possibly due to electrostatic or even hydrodynamic coupling --- the dominant source of change to the swimming velocity comes from solute gradients near the substrate. As mentioned earlier, these are generated by reactions taking place on the swimmer surface and, similar to the way they cause self-propulsion, lead to an effective surface fluid velocity along the wall~\cite{Anderson1989}. This is often referred to as 'slip' velocity, but we do not use this term to avoid confusion with the concept of hydrodynamic surface slip. This effective surface fluid velocity couples back to the swimmer, modifying its net velocity~\cite{Chiang2014, Das2015,uspal2016guiding, popescu2018effective}. In SI Section II, we show that neither purely hydrodynamic coupling~\cite{spagnolie2012hydrodynamics, lintuvuori2016hydrodynamic, Shen2018}, solute confinement~\cite{Popescu2009, Crowdy2013, Uspal2015, Ibrahim2015, Mozaffari2016}, nor reaction-based coupling~\cite{Ebbens2012} can account for the significant wall effect. Instead, as we will show, our observation can be attributed to osmotic coupling~\cite{keh1988electrophoresis, Anderson1989, Chiang2014, Das2015, uspal2016guiding, popescu2018effective}.

The osmotic coupling scales linearly with the slip-velocity parameter $\xi_{w}$,~\textit{i.e.}, the prefactor that converts solute gradients into effective hydrodynamic surface velocities~\cite{Anderson1989}. Ajdari and Bocquet~\cite{Ajdari2006} have shown that for a partial-slip wall the result by Anderson~\cite{Anderson1989} can be generalized to 
\begin{align}
\label{eq:xiw} \xi_{w} &= \left( k_{\mathrm{B}} T / \mu \right) \lambda_{w} \gamma_{w} \left( 1 + b_{w}/\lambda_{w} \right) ,
\end{align}
where slippage is expressed by the slip length $b_{w}$; $b_{w} = 0$ for a no-slip surface and $b_{w} \rightarrow \infty$ for a full-slip surface, respectively. Here, we have introduced $k_{\mathrm{B}}$ Boltzmann's constant, $T$ the temperature, $\mu$ the dynamic viscosity, $\lambda_{w}$ a length scale for the solute-surface interactions, $\gamma_{w}$ a length measuring the solute excess~\cite{Ajdari2006}. For smooth surfaces, as we consider here, the value $\lambda_{w}$ is left relatively unaffected by changes in $\theta$, but $b_{w} \propto (1 + \cos \theta)^{-2}$ and $\gamma_{w} \propto \sqrt{1 + \cos \theta}$~\cite{BocquetAngle2008}. This leads to the following leading-order proportionality of the measured velocity with $\theta$: 
\begin{align}
\label{eq:1}V \propto (1 + \cos \theta)^{-3/2} ,
\end{align}
 which requires that $b_{w}/\lambda_{w} \gg 1$, see also SI Section II for a more in depth discussion on the osmotic coupling based mechanism. We use this quantitative relationship between velocity and contact angle to fit the experimental data presented in Figure~\ref{Fig:mechanism}B. The proportionality factor \textit{A}, which contains all other contributions to the velocity that are slip-independent, is 1.84 $\mu$m/s. The excellent agreement between data and model further corroborates in a quantitative manner the influence of slip.

\begin{figure}
  \centering
 \includegraphics[width=0.75\linewidth=1]{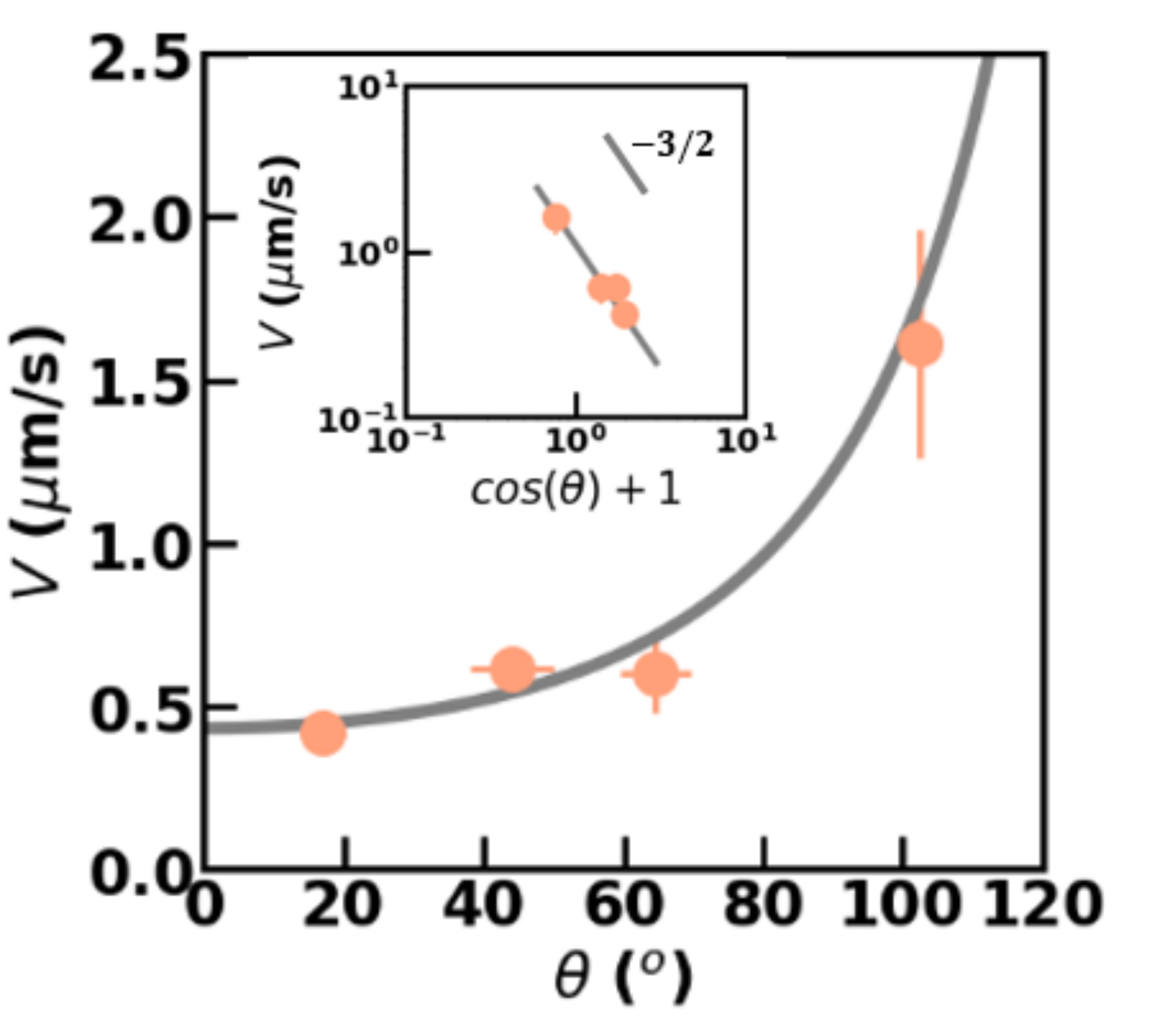}
 \caption{\textbf{Slip dependence of the velocity of colloidal swimmers in salt solution.} Average swimming velocity in 1 mM NaCl as a function of the contact angle $\theta$ and least squares fit (solid line) with $V = A (\cos{\theta}+1)^{-3/2}$ that follows from our model, with $A$ 1.2 $\mu$m/s. The inset shows the data on a log-log scale.}
 \label{Fig:salt}
\end{figure} 

To provide additional support to our hypothesis, we test whether the above dependence persists in the presence of salt. Previous experiments employing 2 $\mu$m PS spheres showed that even 1 mM salt considerably decreases colloid velocities~\cite{Ebbens2014,Brown2014}. Although velocities for similar H$_2$O$_2$ concentration without salt were different above glass, namely around 4 $\mu$m/s~\cite{Ebbens2014} and 18 $\mu$m/s~\cite{Brown2014}, they reduced to 0.45 and 1 $\mu$m/s, respectively, in 1 mM salt. In agreement with these experiments, we find that velocities above different substrates decrease with added salt, see Figure~\ref{Fig:salt}. More importantly, we observe that the velocities still follow the same slip dependence. In salt solution, the proportionality factor \textit{A} is 1.2 $\mu$m/s, i.e. it shows a 33\% decrease compared to the salt-free case. Considering that the influence of salt is complex, potentially affecting zeta potentials, separation, higher-order hydrodynamic moments or possibly more properties including bulk velocities, this decrease is not surprising. However, that the same dependence persists strongly supports the importance of slip, and may provide additional insights into the propulsion mechanism~\cite{Brown2014}. 

We emphasize that other substrate properties besides slip may affect the swimming velocity. As mentioned earlier, lowering the substrate zeta potential has been proposed to increase velocity~\cite{Holterhoff2018}. For completeness thus we measured substrate zeta potentials using a Surface Zeta Potential Cell from Malvern by laser Doppler electrophoresis following Ref.~\cite{Malvern}, see SI Section I E. We find zeta potentials of -38.3~$\pm$ 1.1 mV and -22~$\pm$ 0.9 mV for glass and PDMS, respectively, which is in line with this proposal. However, we find an even lower zeta potential, -11 $\pm$ 5 mV, for hydrophilic PDMS. Based on the low velocity on hydrophilic PDMS, we conclude that the substrate zeta potential is, surprisingly, not the dominant effect. Secondly, an increase in the substrate roughness was shown to increase the velocity~\cite{Holterhoff2018}. We thus performed Atomic Force Microscopy (AFM) measurements, see SI Section I F. The average substrate roughness Ra, is 1.5 and 5 nm, for glass and PDMS, respectively; with Ra denoting the arithmetic mean of the deviations in height from the roughness mean value. Again, this is in line with previous observations. However, hydrophilic PDMS, with a roughness equal to untreated PDMS, featured significantly different velocities. We thus conclude that substrate roughness is also not the dominant effect. 

Finally, we note that the velocity increase on Au-coated surfaces~\cite{Holterhoff2018} may be due to increased surface slip, since contact angles on Au are typically higher than on glass~\cite{Erb1968Gold, Abdelsalam2005Gold}, while the velocity decrease in ~\cite{Wei2018sub} may be due to the hydrophilic polyelectrolyte coatings employed on the glass. Besides, our findings may shed new light on discrepancies in the velocities between previous experiments despite the use of a similar material, coating and solution, as mentioned in the introduction. Even though glass substrates were used, glass can differ in composition, homogeneity and hydrophilicity due to different preparation, coatings, treatment and cleaning methods from the supplier or the researchers themselves, as we have also demonstrated. During contact angle measurements we have noticed that discrepancies of about 10$\degree$ can be found between substrates even within the same batch and/or different parts of the same substrate. AFM indicated that this is likely due to inhomogeneous application or even local absence of coatings applied by the supplier. Thus, similarly to inhomogeneities in colloid properties arising from preparation, inhomogeneities in substrate properties may arise as well. If the coating or treatment that is applied to render the material hydrophilic is inhomogeneous or unstable, locally enhanced substrate slip would persist especially when the underlying material is in principle hydrophobic. Finally, we note that it is possible for substrate slip to be influenced in time by chemically reacting with H$_2$O$_2$; for example, we found a 15$^o$ increase in contact angle for the PS substrate before and after being exposed to H$_2$O$_2$ for several hours.

In conclusion, we show that the velocity of catalytic colloidal swimmers near a wall is influenced by the wall slip boundary condition. This quantitatively follows from theoretical predictions on the basis of an osmotic coupling mechanism, indicating further control and understanding of the behavior of self-propelled particles. Our work points out that wall properties, though often unconsidered, can significantly impact self-propulsion of catalytic swimmers. We expect that the hydrodynamic slip at nearby walls is also relevant for other types of microswimmers.

\acknowledgements
We thank Malvern for providing the Surface Zeta Potential Cell and Sandra Remijn for discussion and help with the substrate zeta potential measurements. We thank Federica Galli for useful discussions and help with the AFM measurements, and Ruben Verweij for help with particle tracking. We are grateful to Aidan Brown, Willem Boon, and Jeroen Rodenburg for fruitful discussions.
J.d.G. thanks NWO for funding through Start-Up Grant 740.018.013 and through association with the EU-FET project NANOPHLOW (766972) within Horizon 2020.
D.J.K. gratefully acknowledges funding from the European Research Council (ERC) under the European Union's Horizon 2020 research and innovation program (grant agreement no. 758383). 

\bibliography{rsc}
\bibliographystyle{unsrtnat}

\end{document}